\pdfoutput=1

\documentclass[12pt, draftclsnofoot, onecolumn]{IEEEtran}

\usepackage{bbm}
\usepackage{amsthm}
\usepackage{amsmath}
\usepackage{amsfonts}
\usepackage{amsmath,amsfonts,amssymb,amsbsy, amsthm,ae,aecompl}
\usepackage{algorithm,algpseudocode}
\usepackage[english]{babel}
\usepackage{bm}
\usepackage{color}
\usepackage[noadjust]{cite}
\usepackage{epsfig}
\usepackage{enumerate}
\usepackage{float} 
\usepackage{fancyhdr}
\usepackage[T1]{fontenc}
\usepackage[acronym,toc,shortcuts]{glossaries}
\usepackage{graphicx, caption, subcaption}
\usepackage{hyperref}
\usepackage{lastpage}
\usepackage{listings}
\usepackage{lipsum}
\usepackage{dsfont}
\usepackage{multirow,tabularx}
\usepackage[normalem]{ulem}
\usepackage{tikz,pgfplots}
\usepackage{times}
\usepackage{verbatim}
\usepackage[all]{xy}
\usepackage{mathtools}
\usepackage{tikz}
\usepackage{tikz-qtree,tikz-qtree-compat}
\usepackage{pgfplots}

\usepackage{lipsum}
\usetikzlibrary{calc}
\usetikzlibrary{patterns}

\usepackage{enumitem}
\graphicspath{{figures/}}

\usetikzlibrary{arrows,shapes}
\newcommand{\E}[1]{{\mathbb E}\left[ #1 \right]}

\newtheorem{Theorem}{Theorem}

\hypersetup{
	colorlinks,%
	citecolor=black,%
	filecolor=black,%
	linkcolor=black,%
	urlcolor=black
}

\pgfplotsset{
	grid style = {
		dash pattern = on 0.025mm off 0.95mm on 0.025mm off 0mm, 
		line cap = round,
		black,
		line width = 0.5pt
	},
	tick label style={font=\small},
	label style={font=\small},
	legend style={font=\footnotesize},
}

\pgfplotscreateplotcyclelist{laneas1}{
	cyan!60!black,		solid, 	every mark/.append style={fill=cyan!60!black},mark=x\\%
	cyan!60!black,		solid, 	every mark/.append style={fill=cyan!60!black},mark=+\\%
	cyan!60!black,		solid, 	every mark/.append style={fill=cyan!60!black},mark=o\\%
	red!80!black,       dashed\\%
	cyan!60!black, 		densely dotted,	every mark/.append style={fill=cyan!80!black},mark=diamond*\\%
}

\ifCLASSINFOpdf
\else
\fi

\begin{document}
%

	\title{Enhancing massive MIMO: A new approach for Uplink training based on heterogeneous coherence times}

		\author{\IEEEauthorblockN{$\text{Salah Eddine Hajri}^*, \text{Mohamad Assaad}^*,
				̃ \text{Maialen Larra\~naga}^*$}\\
				\IEEEauthorblockA{*TCL Chair on 5G, Laboratoire des Signaux et Systemes (L2S, CNRS), CentraleSupelec\\	Gif-sur-Yvette, France\\
					\{Salaheddine.hajri,\;Mohamad.Assaad,\;Maialen.Larranaga\}@centralesupelec.fr}
			
		}


%

\setlength{\columnsep}{0,52cm}
\maketitle
	\begin{abstract}		
		Massive multiple-input multiple-output (MIMO) is one of the key technologies in future generation networks. Owing to their considerable spectral and energy efficiency gains, massive MIMO systems provide the needed performance to cope with the ever increasing wireless capacity demand.
		Nevertheless, the number of scheduled users stays limited in massive MIMO  both in time division duplexing (TDD) and frequency division duplexing (FDD) systems. This is due to the limited coherence time, in TDD systems, and to limited feedback capacity, in FDD mode. 
		In current systems, the time slot duration in TDD mode is the same for all users. This is a suboptimal approach since users are subject to heterogeneous Doppler spreads  and, consequently, different coherence times.
		In this paper, we investigate a massive MIMO system operating in TDD mode in which, the frequency of uplink training differs among users based on their actual channel coherence times. We argue that optimizing  uplink training  by exploiting this diversity can lead to considerable spectral efficiency gain.  We then provide a user scheduling algorithm  that exploits a coherence interval based grouping in order to maximize the achievable weighted sum rate. 
	\end{abstract}
	\pagebreak

	%
	\IEEEpeerreviewmaketitle

	\section{Introduction}
	
	Owing to the exponential increase in data traffic and to the emergence of data-hungry  applications with a large number of connected devices, 5G networks need to be able to cope with huge throughput demand. Several technologies have been considered in order to improve the  wireless networks performance with one noticeable example, namely, massive MIMO Marzetta~\cite{Noncooperative}. A typical  massive MIMO  base station (BS) is equipped with  a large number of  antennas that allow the  spatial  multiplexing  of  a  large  number  of  user devices.
	
	Massive MIMO  systems have been intensively investigated and have shown to have an interesting  potential in improving both the spectral and energy efficiencies of wireless networks. Although promising,  massive MIMO  requires accurate channel state information (CSI) estimates in order to improve the network performance.  Typically, in a TDD system,  CSI is estimated using uplink training with orthogonal pilot sequences. Due to the limited coherence interval,  these sequences are reused which results in  {\it pilot contamination}, see Ashikhmin et al.~\cite{marzetta}.
	Another reason for CSI inaccuracy, which has received less attention, is {\it channel aging}. This phenomenon refers to variation of the channel between the instant when it is learned, via uplink training, and the instant  when it is applied for signal decoding in the uplink and precoding in the downlink. The change of the wireless channel between the BS and the mobile user is mainly due to the mobility of the latter.  The impact of channel aging has been studied in a MIMO system with  coordinated  multi-point transmission/reception (CoMP) in Thiele et al~\cite{aging1}. The authors showed  that  there	is hardly any difference between ideal and delayed feedback when	utilizing channel prediction filters with low mobility.	 The authors in Truong et al.~\cite{aging2} have analyzed the achievable rate in uplink and downlink in the presence of channel aging in massive MIMO. They have proposed a channel prediction scheme that overcomes the performance degradation  due to user mobility. In Papazafeiropoulos et al.~\cite{aging3,aging4}, the effect of channel aging combined with channel prediction has been investigated in scenarios with  regularized ZF precoders and minimum-mean-square-error (MMSE) receivers, respectively. In Kong et al.~\cite{aging5}, lower bounds on the achievable sum rate with channel aging and prediction have been obtained for arbitrary number of BS antennas and  users. Moreover, the power scaling law has been derived for the single cell downlink and multi cell uplink scenarios. The scenario we consider in this paper is different to those proposed in the citations above, in that users in our model are assumed to  have different speeds. This means that the channel aging effect is not the same for all mobile users. More related to our work is Vu et al.~\cite{icc}. The authors in this paper have  proposed  efficient multiuser models for massive MIMO systems that exploit the different user velocities in order to optimize uplink training. The main idea is to allow users with faster changing channels to send uplink training sequences with higher periodicity than users with low velocity.
	
	In this paper, we aim at increasing the  number of  scheduled users per slot without incrementing  training resources.	Users experience different  Doppler spreads, it is therefore sensible to adapt the frequency of their CSI estimation accordingly.  By doing so we reduce the need to estimate the same CSI at each time slot (for slow users). As a result, the  network is able to serve more users while increasing the achievable spectral efficiency. The training method we propose uses an estimate of the Doppler spread or, equivalently, the autocorrelation of the  channel in order to acquire a  channel estimate that can be reused in future slots. An estimate of the channel autocorrelation can easily be obtained by exploiting the cyclic prefix in an OFDM symbol \cite{Est_Doppler}. In our scheme, before uplink training, copilot users are selected based on their channel autocorrelation coefficients. With  copilot users having comparable Doppler spreads, the impact of channel aging on the  achievable spectral efficiency is reduced. This enables the system to reuse the same estimated CSI. While in Vu et al.~\cite{icc}, an equal average achievable rate for all users is considered, we proceed differently by deriving a tight lower  bound on the achievable  rates in a massive MIMO system when using the coherence-time based training briefly introduced above. The derived expression provides more insights into the impact of an adaptive  training procedure where users have heterogeneous CSI delays. Using the derived lower bound, we are able to provide a scheduling optimization algorithm which  achieves a higher weighted sum-rate by exploiting  Doppler spreads diversity.	We formulate a combinatorial optimization problem that determines, for each copilot user group,  whether  it needs to perform uplink training or if it can be processed using outdated  CSI. We then propose a low complexity approximation algorithm that achieves at least a $(4+\alpha)$-approximation of the optimal weighted sum rate.
	
	This paper is organized as follows. We describe the considered system model in Section II.  We discuss the advantages of coherence-time  based  uplink training  in Section III.  We propose a coherence time based scheduling algorithm  in Section IV. Finally, in Section  V, numerical results are presented.  
	\section{System Model And Preliminaries}	
	
	We consider a two-dimensional hexagonal cellular network composed  of $C$ cells with one macro base station (BS) and $K$ users in each cell. Each  BS is equipped with $M$ antennas and each user has a single omni-directional antenna. All users in the network have heterogeneous movement directions and speeds. This means that they experience heterogeneous Doppler spreads and, consequently, different wireless channel correlation across time slots. 
	We assume that the channel of each user can be decomposed as a product of  large and small scale fading coefficients. The wireless channel  from  the $i^{th}$ user in the $b^{th}$ cell to the $j^{th}$ BS, at time slot $t$, is given by:	
	\begin{align}\tag{1}
	& g^{[j]}_{ ib }(t) = \sqrt{\beta^{[j]}_{ ib}} h^{[j]}_{ ib }(t). \nonumber
	\end{align}
	Under flat Rayleigh fading, $h^{[j]}_{ ib } \in \mathbb{C}^{ M \times 1}$ 
	follows a circular-symmetric complex Gaussian  distribution $h^{[j]}_{ ib } \sim CN(0,I_M)$, where $I_M$ is the  identity matrix of size $M\times M$.
	The large-scale fading coefficient ${\beta^{[j]}_{ ib}} \in \mathbb R^+$ depends on the pathloss and the shadowing.  We suppose that the large-scale fading coefficient stays  constant  for  long coherence  blocks and varies much slower than small-scale fading. 
	The time is slotted with a duration of $T_s$ units for each time slot. 	$T_s$ is a system parameter that  depends on the maximum Doppler spread supported by the network \cite{4G}.	We consider a TDD mode where the entire frequency	band is used for downlink and uplink transmissions by all base stations and  users.
	CSI estimation is performed using orthonormal training sequences in the uplink. The same sets of pilot sequences are used in all cells. For $\tau$ scheduled users for uplink training, in each cell,  the orthonormal training sequences $q_i \in \mathbb C^{\tau \times 1} $ ($ q^{\dagger}_i q_j =\delta_{ij} $)  are used.
	In practice, processing delays and user movement causes a variation in the wireless channel between the instants when it is learned and used. This phenomenon is referred to as {\it channel aging}. In order to take into  consideration the autocorrelation of the channel between two time instants, we adopt a time varying model for the wireless channel. 
	We consider   a stationary ergodic Gauss-Markov block fading regular process (or auto regressive model of order 1), where the channel remains constant for a slot duration and changes from slot to slot according to the following equation
	\begin{align}\tag{2}
	& g^{[j]}_{ ib } (t) = \sqrt{\beta^{[j]}_{ ib}} \left(\rho^{[j]}_{ ib } h^{[j]}_{ ib }(t-1)+ \varepsilon^{[j]}_{ ib }(t) \right), \nonumber
	\end{align}
	where  $ \varepsilon^{[j]}_{ ib }(t) $  denotes a  temporally uncorrelated  Gaussian noise with zero mean and  variance  $(1- \rho^{[j]^2}_{ ib }) I_M$.  The temporal correlation parameter of the channel  $\rho^{[j]}_{ ib }$ is given by Jakes~\cite{jakes}: 
	\begin{align}\tag{3}
	& \rho^{[j]}_{ ib } = J_0(2\pi f^{[j]}_{ ib } T_s), \nonumber
	\end{align}
	where $J_0(.)$ is the zeroth-order Bessel function of the first kind and  $f^{[j]}_{ ib }$  represents the maximum Doppler shift of the $i^{th}$ user in cell $b$ with respect to the antennas of the $j^{th}$ BS. In previous works on the channel  aging issue, the same maximum Doppler shift has been considered for all users. In our work,  we adopt a more realistic setting in which, mobile users have different frequency shifts  since we consider different movement velocities and directions. For every user $i,b$ the maximum Doppler shift with respect to the antennas of the $j^{th}$ BS is given by:
	\begin{align}\tag{4}
	& f^{[j]}_{ ib } = \frac{\nu_{ib}f_c}{c} \text{cos}(\theta^{[j]}_{ib}), \nonumber
	\end{align}
	where $\nu_{ib}$ is the velocity of user $i,b$ in meters per seconds, $c= 3 \times 10^8 \text{mps} $ is the speed of light,  $f_c$ is the carrier frequency and $\theta^{[j]}_{ib}$ represents the angle between the movement direction  of the mobile device and the direction of the incident wave. 
	The channel autocorrelation is bounded as, $ 0\leq \left|  \rho^{[j]}_{ ib }\right| \leq 1$. The BS estimates the channels of its associated users thanks to uplink pilot training. In this paper, MMSE channel estimation is used. 
	The  wireless channel of user $i,b$, at slot $t$, can be  written  as follows:
	\begin{align}\tag{5}
	& g^{[j]}_{ ib } (t) = \rho_{ ib } \hat{g}^{[j]}_{ ib } (t-1) + \rho_{ ib } \tilde{g}^{[j]}_{ ib } (t-1) +   \sqrt{\beta^{[j]}_{ ib}} \varepsilon^{[j]}_{ ib }(t), \nonumber
	\end{align}
	where $\hat{g}^{[j]}_{ ib } (t-1) $ is the MMSE channel estimate at time $t-1$ which has a  $\hat{g}^{[j]}_{ ib } (t-1) \sim CN \left(0,\frac{ \beta^{[j]^2}_{ ib} }{\frac{1}{P_{p}}+\sum_{l, l\neq b}^{C} \beta^{[j]}_{ il} } I_M \right) $ distribution with $P_{p}$  the uplink pilot  transmit power. The channel estimation error at time $t-1$  is denoted $\tilde{g}^{[j]}_{ ib } (t-1)$ with  $\tilde{g}^{[j]}_{ ib } (t-1) \sim CN \left(0,\left(\beta^{[j]}_{ ib}- \frac{ \beta^{[j]^2}_{ ib} }{\frac{1}{P_{p}}+\sum_{l, l\neq b}^{C} \beta^{[j]}_{ il} } \right) I_M \right) $. 
	Note that $\hat{g}^{[j]}_{ ib } (t-1) $ and $\tilde{g}^{[j]}_{ ib } (t-1)$ are mutually independent. 
	\section{A new coherence time based approach for  user Scheduling In Massive MIMO Systems}
	The current  massive MIMO models consider a fixed coherence  interval $T_s$ for all  users in the network.  The duration of $T_s$ is based on the maximum Doppler spread that the network can handle \cite{4G}.
	Sidestepping the important fact that, users experience heterogeneous Doppler spreads and, consequently, different coherence intervals, can only result in  suboptimal exploitation of the available resources.	In fact,  requiring users to  perform  uplink CSI estimation at the same frequency without taking into  consideration  their actual channels   autocorrelation, results in  a useless redundancy of information.
	If we take into consideration the  channel aging effect between time slots, we notice that, whenever the variation of channel  is not important, we may  reuse outdated CSI estimates in order to reduce the need for further training.  
	In practice, this  means that the network is not required  to  estimate the channels of all scheduled users at each time slot. A more appropriate approach is to  estimate the CSI of scheduled users whenever needed, as a function of  its actual  coherence interval. The idea of reusing an already estimated information can be quite tempting. It frees part of the training resources. These can be used for data transmission or to schedule more users. In both  cases, we can  observe  an increase in the  achievable  spectral efficiency. Nevertheless, allowing the network  to reuse the  CSI from past slots can not provide the needed capacity increase if no prior conditions on  user selection is imposed. 
	In this section, we investigate the achievable rate with outdated CSI. We also derive an important  condition which  ensures that a coherence-time based training scheme is able to provide a considerable increase in the network performance. 
	
	\subsection{An adaptive coherence time based user scheduling}

\begin{figure}[h!]
		\centering
		\includegraphics[width=10cm,height=3.5cm]{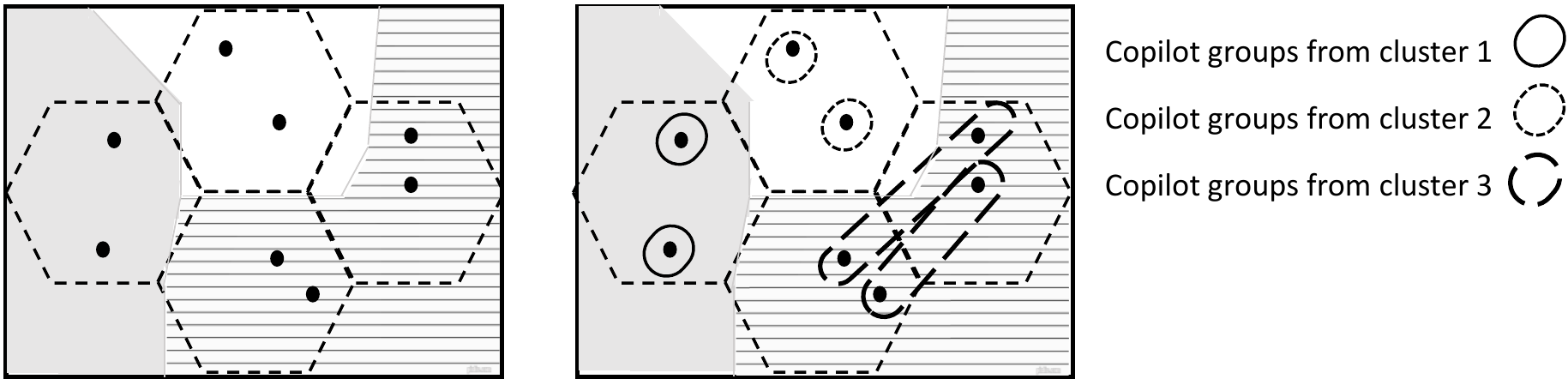}
		\caption{Example of the adaptive coherence-time based user scheduling procedure with $C=4, K=2, N_g=6, N_c=3$. {\bf Left:} illustration of Step 2). {\bf Right:} illustration of Step 3).}\label{System Model}
	\end{figure}

	We consider a massive MIMO system,  where the network  is not  required to  estimate the  CSI of all  scheduled users at  each time slot. The only requirement is that copilot users have the same CSI delay. This guarantees a proportional aging effect for copilot users. The latter provides a performance gain that will be explained later in this section. For all users in the same copilot group to have the same CSI delay, we consider that all users in the same copilot group $\lambda_g$ for $g=1,...,N_g$  are either scheduled for uplink training synchronously, and they have the same CSI delay   $d_g, g=1,...,N_g$ or are not scheduled at all. 
	At each time slot, the network schedules all covered users for data transmission, and a maximum of $\tau<N_g$ copilot groups are allowed to perform uplink training. The rest will be using the last estimated version of their CSI.   
	Instead of requiring the scheduled users to perform uplink  training with  the same periodicity, we propose a new Time Division Duplexing protocol where, the training frequency depends on the  actual  coherence time. The proposed adaptive coherence-time based user scheduling procedure goes as follows (see also Figure~\ref{System Model}).
	
	\begin{enumerate}
		\item In the beginning of each  large-scale coherence block, the network estimates the large scale fading  and  channel  autocorrelation coefficients, i.e., $\beta_{ib}^{[j]}$ and $\rho_{ib}^{[j]}$ for all $i=1,\ldots,K$, and $b,j=1,\ldots,C$.
		\item Next, the network  constructs $N_c$ user clusters, based on the autocorrelation coefficients, using the $K$-mean algorithm with  $K=N_c$, see Young et al~\cite{mean}. Each cluster will  be characterized by an average autocorrelation coefficient or, equivalently, an average Doppler spread. Defining the number of clusters  $N_c$  is of paramount importance. In this work, we choose to define $N_c$ as
		\begin{align}\tag{6}
		\nonumber& N_c= \lceil\frac{T_{max}}{T_{s}}\rceil,\nonumber
		\end{align}
		where $T_{max}$ represents the maximum coherence time experienced by the users. Defining $N_c$ as in $(6)$ is motivated by the need to have an average coherence slot per cluster  that approach a multiple of  $T_{s}$. Doing so guarantees a proper definition of the  periodicity  of CSI  estimation as a function of $T_{s}$.
		\item Next, all users in the network ($K$ per cell) will be allocated to $N_g$ copilot user groups. Each  group contains at maximum $C$ users from the same channel autocorrelation  cluster and from different cells. 
		\item The network then  schedules $\tau$ copilot users for uplink training synchronously. Selecting the users to  be scheduled for  uplink training can be  done using various types of scheduling algorithms depending on the desired performance criteria. The algorithm we propose is to be found in Section~IV. 
	\end{enumerate}
	

	\subsection{Spectral efficiency with  outdated CSI}
	We now investigate the achievable spectral efficiency when the aforementioned  procedure is applied. For the sake of analytical traceability, we consider that all copilot groups contain exactly $C$ users.	We, henceforth, refer to each user  by its copilot group and serving BS indexes $g,l$. 
	During uplink data transmission, the  $j^{th}$ BS receives the  following  data signal at time slot $t$:						
	\begin{align}\tag{7}
	Y^{[j]}_u(t) \nonumber&=  \sum_{l=1}^{C} \sum_{g=1}^{N_g} \sqrt{P_u} g^{[j]}_{ gl } (t) S_{gl} +W_u(t), \nonumber
	\end{align}
	where $P_u$ denotes  the reverse link  transmit power,  $W_u(t) \sim CN(0,I_M)$ is the additive noise and $S_{gl}$ denotes the  uplink signal of the user from copilot group $g$ in cell $l$.
	In what follows we derive lower bounds on the achievable sum rate with a matched filter receiver when  outdated CSI is used.  
	At the reception, the $j^{th}$ BS applies a matched filter receiver based on the  latest version of estimated CSI for the scheduled users. The resulting average achievable sum rate in the system is given in the following theorem. The proof can be found in Appendix~A  in \cite{long_version}.
	
	\begin{Theorem}
		The network serves $N_g$ copilot groups, $\tau$  of which are scheduled for uplink training.  Using a matched filter receiver based on the latest available CSI estimates for each group, the average achievable sum rate $R_u $ in the uplink is lower bounded by:
		\begin{align}\tag{8}
		\nonumber& R_u \geq \sum_{l=1}^{C} \sum_{g=1}^{N_g}
		\left( 1-\frac{\tau}{T_s}\right) \text{log} \left(1+ \frac{(M-1)\beta^{[l]^2}_{gl} \rho^{[l]^{2 d_g}}_{ gl } }{(M-1) \times I^p_{gl} +  I^n_{gl}  } \right),\nonumber	 
		\end{align}
		where $d_g, g=1...N_g$ represents the CSI delays of users using the same pilot sequence. $I^p_{gl}$ and $I^n_{gl}$ are given by:
		\begin{align}
		\tag{9}\nonumber& I^p_{gl} = \sum_{ c\neq l}^{C} \rho^{[l]^{2 d_g}}_{ gc } \beta^{[l]^2}_{gc}, \\\nonumber
		\nonumber& I^n_{gl}= (\sum_{c=1}^{C} \sum_{k\neq g}^{} \beta^{[l]}_{kc} +  \sum_{ c=1}^{C} ( \beta^{[l]}_{gc} -    \rho^{[l]^{2 d_g}}_{ gc }\frac{ \beta^{[l]^2}_{gc}}{\frac{1}{P_{p}}+\sum_{b=1}^{C}  \beta^{[l]}_{gb} })\\\nonumber 
		\tag{10}\nonumber& + \frac{1}{P_u})\times (\frac{1}{P_{p}}+ \sum_{b=1}^{C} \beta^{[l]}_{gb}  ). \nonumber	 
		\end{align}	
	\end{Theorem}
	Equation $(8)$ provides further insights into the evolution of the achievable sum rate as a function of the CSI time offset.  The achievable sum rate for a given copilot user group decreases as the time offset increases since the correlation between the estimated CSI and the actual channel  fades over time. 
	In order to investigate the potential gain that the proposed approach can provide, we compare it with a reference model in which all of the $N_g$ copilot groups take part in the uplink training at each time slot, as in  classical TDD protocols. This comparison is done in the next theorem where we consider the asymptotic regime (as $M$ grows large). The proof can be found in Appendix~B  in \cite{long_version}.

	\begin{Theorem}
		In the asymptotic regime, the proposed training approach enables to improve the achievable spectral efficiency of each user if the following condition is satisfied $\forall \; \lambda_g, g=1,...,N_g$:
		\begin{align}\tag{11}
		\nonumber& \left( \frac{\bar{\rho}^{[{min}]^2}_{g}}{\bar{\rho}^{[{max}]^2}_{{g}} }\right)^{d_g} \geq \frac{ \left( 1+\text{ SINR}^{[\infty]}_{g,l}\right)^{\frac{T_s-N_g}{T_s-\tau}}-1}{\text{ SINR}^{[\infty]}_{g,l}},  \nonumber
		\end{align}
		with
		\begin{align}\tag{12}
		\nonumber& \text{ SINR}^{[\infty]}_{g,l}= \frac{\beta^{[l]^2}_{gl}  }{	\sum_{ b\neq l}^{}  \beta^{[l]^2}_{gb}},  \nonumber
		\end{align}	
		and $\bar{\rho}^{[{min}]}_{g}$ and $\bar{\rho}^{[{max}]}_{{g}}$ the minimum and maximum channel autocorrelation  coefficients in  group $g$, respectively. 
		
	\end{Theorem}
	
	Condition~$(11)$ ensures that the achievable  spectral  efficiency of  all users is improved when outdated CSI is  used. We can see that the gain in spectral efficiency is maintained as long as the SINR degradation over time is compensated by  the spared resources from  uplink training. 
	It also  implies that the ratio between the maximum  and minimum  autocorrelation should be  bounded. The bound becomes tighter as the  allowed CSI delay  increases. This condition is quite intuitive. It means that copilot users are required to  experience comparable channel  aging effects in order to  be  able  to  improve the  achievable  spectral  efficiency  using  less  uplink training. In order for  Condition~(11) to be satisfied we need users from the same copilot group to have equivalent coherence times and, consequently, they proceed to uplink training with the same periodicity. This is the reason why in  Step $2)$ of the scheme proposed in Section~III.A we group the mobile users according to  channel  autocorrelation coefficients.
	Next, we  provide an efficient  user  scheduling method for uplink training. It exploits the autocorrelation based grouping in order to improve the achievable weighted sum rate. We consider that all copilot groups are active during data transmission. The network is required to select which groups will refresh their CSI by scheduling them for  training.
	
	\section{Improving the achievable weighted sum rate} 
	
	Assuming  a constant  coherence interval for all users results in a suboptimal  resources exploitation since in practice, users  experience  heterogeneous Doppler frequencies.  We showed, in Theorem  $2$, that  allowing the  network  to  use outdated CSI estimates can  actually improve the achievable spectral  efficiency. In this  section, we study  the impact  of  the proposed adaptive training scheme on the achievable weighted sum rate. In this  case, scheduling users for uplink  training does not depend solely on the  estimated CSI but  also on the users traffic pattern.   
	In order to  take into consideration  the QoS, we consider optimizing the weighted sum of rates. The weights enable to introduce fairness in the scheduling procedure \cite{Stability1}.  We denote the weights associated with  the users by	$w_{gl},  \; \forall g=1,...,N_g , \; l =1,...,C   $. 
	
	The aim here is to  schedule copilot groups for uplink training in order to maximize the achievable weighted sum rate. As in previous sections we assume that all users  are scheduled for data transmission and $\tau<N_g$ copilot groups are selected to update their CSI.
	Grouping users based on their  autocorrelation coefficients simplifies the scheduling problem.  Instead of deciding, at each time slot, to which user each pilot sequence is going to be allocated, decisions are made on predefined collections of copilot users. We define the vector $Y=(y_1,\ldots,y_{N_g})$,  with $y_{g}, \forall g=1,...,N_g$  given by
	\begin{equation}\tag{13}
	\begin{aligned}
	& y_{g}=
	\begin{cases}
	1 & \mbox{if  group $g$ is scheduled for training,}  \\
	0 & \mbox{otherwise. } 
	\end{cases}
	\end{aligned}
	\end{equation}
	The copilot group scheduling problem when outdated CSI use is permitted, is formulated as follow
	\begin{align}\tag{14}
	\underset{Y}{\text{max}}&  (1-\frac{\sum_{g=1}^{N_g} y_g}{T_s})\sum_{l=1}^{C} \sum_{g=1}^{N_g} w_{gl}R_{gl}(\vec d,Y), \\\nonumber\tag{14a}
	\text{subject to}& \sum_{g=1}^{N_g} y_{g} \leq \tau, \nonumber
	\end{align}
	where $R_{gl}(\vec d,Y)$ is the lower bound on the achievable sum rate for the user in copilot group $g$ and cell $l$ (see Eq. (8)), and $\vec d$ is the delay vector.
	Here, $(14a)$ captures the fact that the number of scheduled users for uplink training in each cell, is at maximum equal to $\tau$. 
	In the following theorem, we show that problem $(14)$ is equivalent to maximizing a submodular  set function subject to a cardinality constraint. This formulation  enables to apply a low complexity algorithm that solves the problem. The proof can be found in Appendix~C in \cite{long_version}.
	\begin{Theorem}
		Problem $(14)$ is equivalent to maximizing a submodular  set function with a cardinality constraint. 
	\end{Theorem}
	
	In  order to  solve problem  $(14)$, we  use the Submod-Max-Cardinality algorithm in Gupta et al. \cite{card_sub}. This algorithm starts by running the commonly used greedy approach, see Nemhauser et al.~\cite{greed}. This provides a first candidate solution $S_1$.  The approximation  algorithm  of Feige et al. \cite{local_search} is then used on $S_1$ in order to obtain a second candidate solution $S_2$. The same steps are repeated on the remaining set of copilot groups that were not selected in $S_1$.  The algorithm returns then the best found solution.
	When used to solve a non-monotone submodular function maximization subject to  a cardinality  constraint, Submod-Max-Cardinality yields  a $(4+\alpha)$-approximation of the optimal solution \cite{card_sub}.  The detailed algorithm can be found in Section~IV \cite{long_version}

	\section{Numerical Results} 
	In this section we provide some numerical results demonstrating the performance of the proposed training/copilot group scheduling scheme. Our training procedure is compared with  a massive MIMO system operating according to the  classical TDD protocol, considered as a reference model.    
	We consider an hexagonal cell network  with $C=7$ cells. Each cell has a $1.5\; \text{Km}$ radius. The mobile users are  located uniformly at random in each cell and we assume that no user is  closer  than $r_0= 10 \;\text{m}$ to its serving BS.
	We consider heterogeneous movement speeds and directions. User  speeds are randomly generated  in the interval $\left[20\;\text{Km/h}\;,\;80 \;\text{Km/h}\right] $. User movement direction, which is defined by   the angle between the movement direction  of the mobile device and the direction of the incident wave, is also randomly generated over the interval $\left[0\;,\;2 \pi \right]$.
	We consider a path-loss exponent $\sigma=3.5$ and a relatively short coherence block of $T_s=200$ symbols is considered. We suppose 
	a $ 200 \; Mhz$  bandwidth \cite{5G}.
	The spectral efficiency is measured in $\text{bits/s/hz}$.
	The CSI  delays for each   copilot group $d_g, g=1,...,N_g$   are selected randomly over the interval $[0,d_{max}]$. 
	
		\begin{figure}[h!]
			\centering
			\includegraphics[width=11cm,height=9cm]{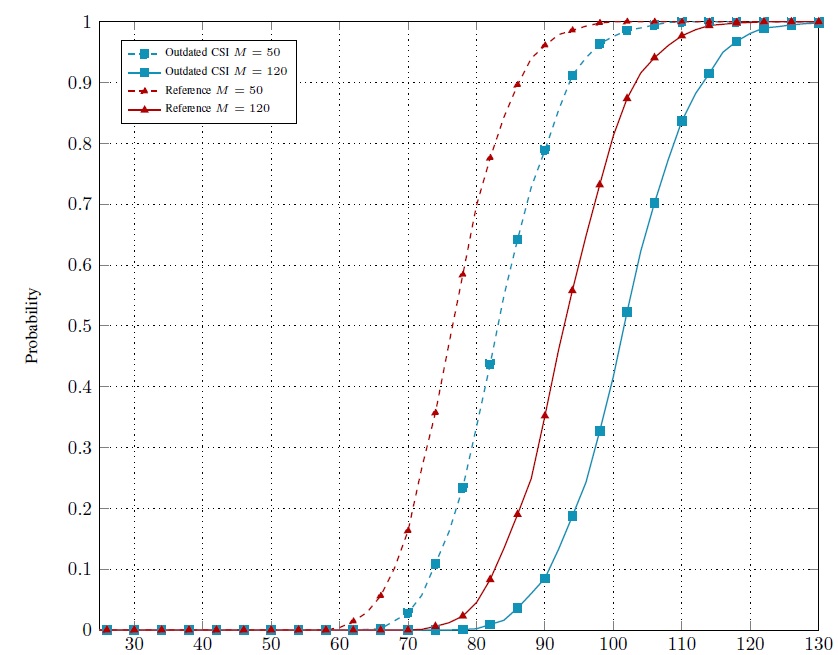}
			\captionsetup{font=footnotesize}
			\caption{ Comparison of the  CDFs of spectral efficiency ($N_g=30,\; \tau=15, \; d_{max}=2 $ )}
		\end{figure}

	In	Figure 2,  we illustrate a comparison of the CDFs of the  achievable spectral efficiency between the reference  model and  the proposed outdated CSI based training scheme for different numbers of antennas at the BS. Figure 2 shows that, for $50$ receive antennas at the BS, the  proposed training scheme achieves $5\%$-outage  rate around $71\; \;\text{bit/s/hz}$.  This performance represents a gain of $ 5 \;\text{bit/s/hz}$ or, equivalently, $1000\;\text{Mbit/s}$ for a system bandwidth of $ 200 \; Mhz$ \cite{5G},  compared with the reference model that achieves $5\%$ outage  rate around $66\; \;\text{bit/s/hz}$.  For    $120$  antennas at the BS,
	the gain in the  $5\%$-outage  rate  when, outdated CSI is  used,  attains $1400\;\text{Mbit/s}$. In fact, while the proposed training scheme achieves $5\%$-outage  rate around $87\; \;\text{bit/s/hz}$, the reference model
	achieves the same outage rate around $80\; \;\text{bit/s/hz}$.
	This gain  comes from  the spared resources during uplink training. In fact, using outdated CSI enables to  increase the number of channel  uses that are allocated to  data transmission.

		\begin{figure}[h!]
			\centering
			\includegraphics[width=11cm,height=9cm]{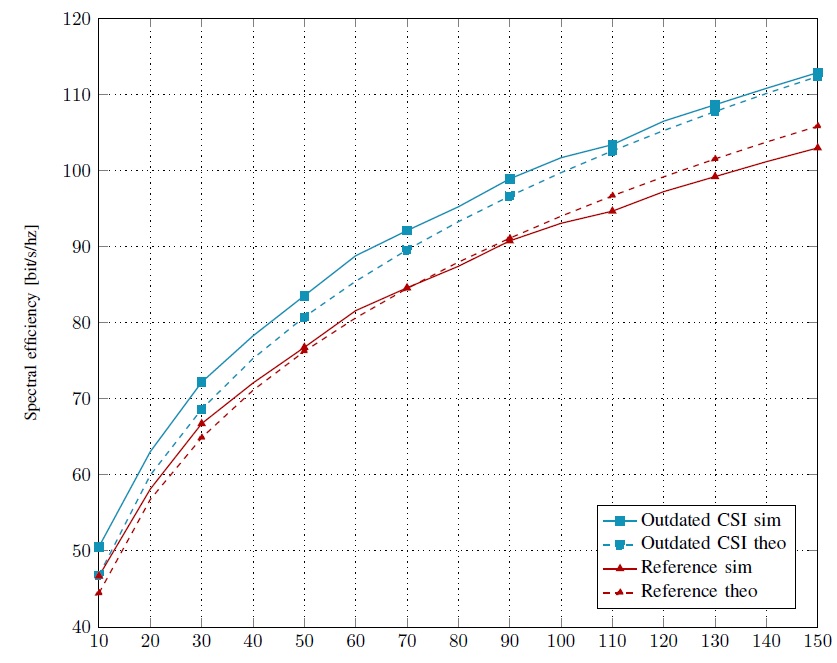}
			\captionsetup{font=footnotesize}
			\caption{Spectral efficiency for varying values of $M$ ($N_g=30,\; \tau=15, \; d_{max}=2  $)}
		\end{figure}

	In	Figure 3, we illustrate the relation between the  achievable spectral efficiency and the number of BS antennas. We compare the lower bound in Theorem $1$ with a simulated curve. Figure 3 shows that the derived is tight for a wide range of $M$. This figure also  shows that, when compared with the  reference model, the  proposed training scheme  increases the  achievable spectral efficiency by $ 6.8 \;\text{bit/s/hz}$ for $M=50$. This is equivalent to a rate gain of $1360\;\text{Mbit/s}$ for a system bandwidth of $ 200 \; Mhz$ \cite{5G}. This gain  attains $ 9.9\; \text{bit/s/hz}$ or, equivalently, $1980\;\text{Mbit/s}$ for $M=150$.

		\begin{figure}[h!]
			\centering
			\includegraphics[width=11cm,height=9cm]{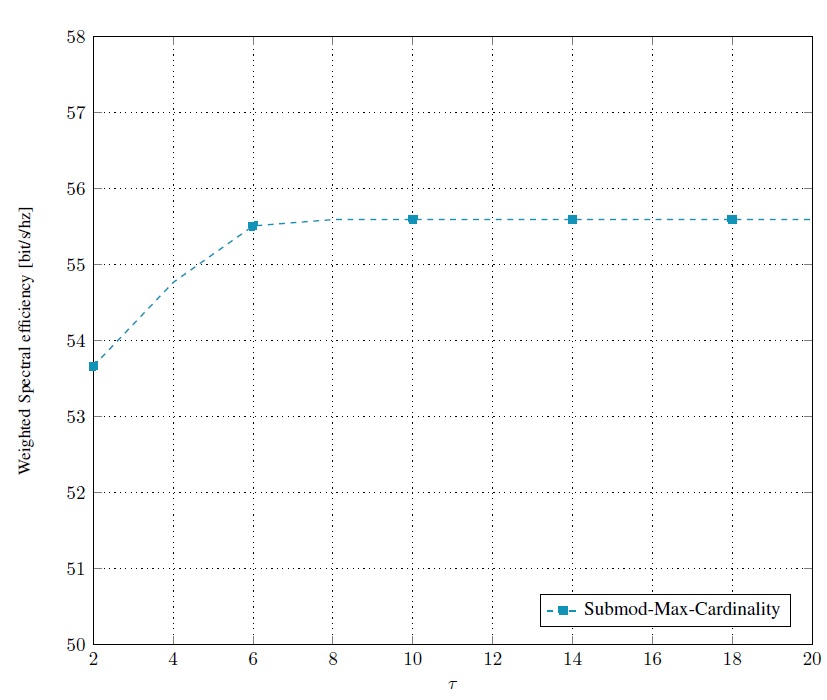}
			\captionsetup{font=footnotesize}
		\caption{Weighted Spectral efficiency  for varying value of $\tau$ ($N_g=20,\; M=100, \; d_{max}=3 $ ) }
		\end{figure}

	In	Figure 4, we illustrate the performance of the proposed training/copilot group scheduling scheme. The weights  $w_{gl},  \; \forall g=1,...,N_g , \; l =1,...,C   $ are randomly generated in the interval $ [0,1]$.
	Figure 4 shows that the optimal weighted sum rate  saturate at $\tau=8$. This means that the optimal number of copilot groups scheduled for uplink training is $8$ in this setting. Consequently, scheduling more users for training will not result in a weighted sum rate gain. This actually concords with the suggestion that not all users who are active for data transmission are required to participate in uplink training. Consequently, an adaptive training scheme that takes into consideration CSI aging is justified.

	\pagebreak
	\section{Conclusion}
	In this paper, we studied the achievable sum rate in a massive MIMO system with a matched filter receiver taking into account the channel aging effect.
	We proposed a training scheme in which,  we deliberately utilize outdated CSI estimates in order to optimize uplink training.
	The derived results show that,  although outdated CSI  degrades the achievable channel gain, we are able to achieve better spectral efficiency  when training frequency is adapted to the channel's autocorrelation.
	We show that, grouping the users in autocorrelation based clusters and optimizing their scheduling accordingly, provides a substantial increase  in the achievable weighted sum rate. Future works will include investigating more sophisticated algorithms that enable to further leverage heterogeneous coherence times.

		\section{Appendix}

				\subsection{Proof of Theorem 1}
				
				The network serves $N_g$ copilot groups, $\tau$  of which are scheduled for uplink training. At the reception, each  BS uses a matched filter receiver that is based on the latest available CSI estimates. BS  $l$ detects the signal of user $g$ in cell $l$ by applying the following filter
				\begin{align*}\tag{17}
				& u_{gl}(t)= \frac{\hat{g}^{[l]}_{ gl } (t-d_{g}) }{\lVert \hat{g}^{[l]}_{ gl } (t-d_{g})  \rVert}, t\geq d_{g}, 
				\end{align*}
				where $\hat{g}^{[l]}_{ gl } (t-d_{g})$ denotes the latest available  CSI estimate  for user $g$ in cell $l$.
				Consequently, the detected signal  of user $g$ in cell $l$ is given by the following
				\begin{align}\tag{18}
				u_{gl}^\dagger(t) \frac{Y^{[l]}_u(t)}{\sqrt{P_u}}\nonumber&= u_{gl}^\dagger(t)(\sum_{k=1}^{N_g} \sum_{c=1}^{C}   g^{[l]}_{ kc } (t) S_{kc} +\frac{W_u(t)}{\sqrt{P_u}} )  \nonumber\\
				&=u_{gl}^\dagger(t)(g_{gl}^{[l]}(t)S_{gl}+\sum_{c \neq l}^Cg_{gc}^{[l]}(t)S_{gc}+\sum_{k\neq g}^{N_g}\sum_{c=1}^Cg_{kc}^{[l]}(t)S_{kc}+\frac{W_u(t)}{\sqrt{P_u}})\nonumber\\
				&=u_{gl}^\dagger(t)((\rho^{[l]}_{ gl })^{d_{g}} \hat{g}_{gl}^{[l]}(t-d_g)S_{gl}+\sum_{c \neq l}^C (\rho^{[l]}_{ gc })^{d_{g}}\hat{g}_{gc}^{[l]}(t-d_g)S_{gc}+\sum_{c=1}^C(\rho^{[l]}_{ gc })^{d_{g}}\tilde{g}_{gc}^{[l]}(t-d_g)S_{gc}\nonumber\\
				&+\sum_{c=1}^C\sum_{j=0}^{d_g-1} (\rho^{[l]}_{ gc })^{j} \sqrt{\beta_{gc}^{[l]}} \varepsilon_{gc}^{[l]}(t-j)S_{gc}+\sum_{k\neq g}^{N_g}\sum_{c=1}^Cg_{kc}^{[l]}(t)S_{kc}+\frac{W_u(t)}{\sqrt{P_u}})\nonumber\\
				&=u_{il}^\dagger(t) (I_1(t)+I_2(t)+I_3(t))\nonumber,
				\end{align}
				with
				\begin{align*}
				I_1(t) &= (\rho^{[l]}_{ gl })^{d_{g}}\hat{g}_{gl}^{[l]}(t-d_g)S_{gl},\\
				I_2(t) &= \sum_{c \neq l}^C  (\rho^{[l]}_{ gc })^{d_{g}} \hat{g}_{gc}^{[l]}(t-d_g)S_{gc},\\
				I_3(t) &=  \sum_{c=1}^C(\rho^{[l]}_{ gc })^{d_{g}}\tilde{g}_{gc}^{[l]}(t-d_g)S_{gc}+\sum_{c=1}^C\sum_{j=0}^{d_g-1} (\rho^{[l]}_{ gc })^{j} \sqrt{\beta_{gc}^{[l]}} \varepsilon_{gc}^{[l]}(t-j)S_{gc}+\sum_{k\neq g}^{N_g}\sum_{c=1}^Cg_{kc}^{[l]}(t)S_{kc}+\frac{W_u(t)}{\sqrt{P_u}} \\
				\end{align*}
				The third equality in Equation $(18)$ follows from the fact that $g_{ic}^{[l]}(t)=\sqrt{\beta_{ic}^{[l]}}h_{ic}^{[l]}(t)$, $h_{ic}^{[l]}(t)=\rho_{ic}^{[l]}h_{ic}^{[l]}(t-1)+\varepsilon_{ic}^{[l]}(t)$ for all $t$ and $g_{ic}^{[l]}(t)=\hat g_{ic}^{[l]}(t)+\tilde g_{ic}^{[l]}(t)$ for all $t$.
				
				We note that $I_1(\cdot)$ refers to  the useful  signal,  $I_2(\cdot)$ represents the impact of pilot contamination and $I_3(\cdot)$ regroups the impact of the  white noise, channel estimation error, non correlated interference due to users with different pilot sequences and the impact of channel aging.  The instant spectral efficiency attained by user $g$ in cell $l$ is:
				\begin{align*}\tag{19}
				R_{g,l} = \left(1-\frac{\tau}{T_s}\right)\log\left(1+\frac{|u_{gl}^\dagger(t) I_1(t)|^2}{|u_{gl}^\dagger(t) I_2(t)|^2+|u_{gl}^\dagger(t) I_3(t)|^2}\right).
				\end{align*}
				We now define $\overline R_{g,l}$ to be the average achievable sum rate of user $g$ in cell $l$, namely,
				\begin{align*}\tag{20}
				\overline R_{g,l}&=\mathbb{E}\left(\left(1-\frac{\tau}{T_s}\right)\log\left(1+\frac{|u_{gl}^\dagger(t) I_1(t)|^2}{|u_{gl}^\dagger(t) I_2(t)|^2+|u_{gl}^\dagger(t) I_3(t)|^2}\right)\right)\\
				&=\mathbb{E}\left(\mathbb{E}\left(\left(1-\frac{\tau}{T_s}\right)\log\left(1+\frac{|u_{gl}^\dagger(t) I_1(t)|^2}{|u_{gl}^\dagger(t) I_2(t)|^2+|u_{gl}^\dagger(t) I_3(t)|^2}\right)\bigg|\hat g_{gl}^{[l]}(t-d_{g})\right)\right),
				\end{align*}
				the last equality follows from the law of total expectation. Let us define $\overline R_{g,l}^0$ such that
				\begin{align*}\tag{21}
				\overline R_{g,l}^0 = \mathbb{E}\left(\left(1-\frac{\tau}{T_s}\right)\log\left(1+\frac{|u_{gl}^\dagger(t) I_1(t)|^2}{|u_{gl}^\dagger(t) I_2(t)|^2+|u_{gl}^\dagger(t) I_3(t)|^2}\right)\bigg|\hat g_{gl}^{[l]}(t-d_{g})\right),
				\end{align*}
				therefore, 
				\begin{align}\tag{22}
				\overline R_{g,l} = \mathbb{E}(\overline R_{g,l}^0 ).
				\end{align}
				Based on the convexity of $\text{log}(1 +\frac{1}{x+a})$, and Jensen's  inequality  the following inequality can be obtained
				\begin{align}\tag{23}
				\overline R^0_{g,l}\geq\left(1-\frac{\tau}{T_s}\right)\log\left(1+\frac{|u_{gl}^\dagger(t) (\rho_{gl}^{[l]})^{d_{g}}\hat g_{gl}^{[l]}(t-d_{g})|^2}{\mathbb{E}(|u_{gl}^\dagger(t) I_2(t)|^2|\hat g_{gl}^{[l]}(t-d_{g}))+\mathbb{E}(|u_{gl}^\dagger(t) I_3(t)|^2|\hat g_{gl}^{[l]}(t-d_{g}))}\right),
				\end{align}
				since 
				\begin{align}\tag{24}
				\mathbb{E}(|u_{gl}^\dagger(t)I_1(t)|^2|\hat g_{gl}^{[l]}(t-d_{g}))=|u_{gl}^\dagger(t)(\rho_{gl}^{[l]})^{d_{g}}\hat g_{gl}^{[l]}(t-d_{g})|^2,
				\end{align}
				by the property $\mathbb{E}(f(Z)|Z)=f(Z)$ for a random variable $Z$.
				We now aim at computing 
				$\mathbb{E}(|u_{gl}^\dagger(t) I_j(t)|^2|\hat g_{gl}^{[l]}(t-d_{g})) \hbox{ for } j=2,3.$  
				In order to do so, we are first going to obtain an alternative expression for $I_2(t)$, that is, 
				\begin{align*}\tag{25}
				I_2(t) &= \sum_{c \neq l}^C\hat{g}_{gc}^{[l]}(t-d_g)S_{gc}= \hat{g}_{gl}^{[l]}(t-d_g) \sum_{c \neq l}^C \frac{\beta_{gc}^{[l]}}{\beta_{gl}^{[l]}}  S_{gc},\\
				\end{align*}
				since $\hat{g}_{gc}^{[l]}(t-d_g)=\hat{g}_{gl}^{[l]}(t-d_g)\frac{\beta_{gc}^{[l]}}{\beta_{gl}^{[l]}} $.	Therefore, $I_2(t)$ and $\hat{g}_{gl}^{[l]}(t-d_g)$ are correlated. Consequently, we obtain
				\begin{align}\tag{26}
				&\E{|u_{gl}^\dagger(t) I_2(t)|^2|\hat g_{gl}^{[l]}(t-d_{g})} =   \left| u_{gl}^\dagger(t)  \hat{g}^{[l]}_{ gl } (t-d_{g})    \right|^2    \sum_{c \neq l}^C (\rho^{[l]}_{ gc })^{2 d_{g}} \frac{\beta^{[l]^2}_{gc}}{\beta^{[l]^2}_{gl}}.
				\end{align}
				We will now compute $\mathbb{E}(|u_{gl}^\dagger(t) I_3(t)|^2|\hat g_{gl}^{[l]}(t-d_{g}))$. First note that, $I_3(t)$ is independent of $\hat g_{gl}^{[l]}(t-d_{g})$ and since $u_{gl}^\dagger(t)$ has unit norm, we have that $\mathbb{E}(|u_{gl}^\dagger(t) I_3(t)|^2|\hat g_{gl}^{[l]}(t-d_{g}))=\mathbb{E}(|I_3(t)|^2)$, therefore we obtain 
				\begin{align}\tag{27}
			\nonumber	\mathbb{E}(|I_3(t)|^2 )&=\mathbb{E}\bigg(|\sum_{c=1}^C (\rho_{gc}^{[l]})^{d_g}\tilde{g}_{gc}^{[l]}(t-d_g)S_{gc}+\sum_{c=1}^C\sum_{j=0}^{d_g-1} \sqrt{\beta_{gc}^{[l]}}(\rho_{gc}^{[l]})^j\varepsilon_{gc}^{[l]}(t-j)S_{gc}\\\nonumber
				&+\sum_{k\neq g}^{N_g}\sum_{c=1}^Cg_{kc}^{[l]}(t)S_{kc}+\frac{W_u(t)}{\sqrt{P_u}}|^2\bigg)\nonumber\\
			\nonumber	&=\mathbb{E}\bigg(\sum_{c=1}^C |(\rho_{gc}^{[l]})^{d_g}\tilde{g}_{gc}^{[l]}(t-d_g)|^2+\sum_{c=1}^C\sum_{j=0}^{d_g-1} |\sqrt{\beta_{gc}^{[l]}}(\rho_{gc}^{[l]})^j\varepsilon_{gc}^{[l]}(t-j)|^2\\\nonumber
			\nonumber	&+\sum_{k\neq g}^{N_g}\sum_{c=1}^C|g_{kc}^{[l]}(t)|^2+|\frac{W_u(t)}{\sqrt{P_u}}|^2\bigg),\nonumber
				\end{align}
				where the last equality follows from noting the following four properties; (i) $S_{kc}\cdot S_{ic'}=0$ for all $k\neq i$ and all $c,c'\in\{0,\ldots,C\}$, (ii) $\mathbb{E}(Z W_u(t))=\mathbb{E}(Z)\mathbb{E}(W_u(t))=0$ for all random variables $Z$ that are independent of $W_u(t)$ (zero mean complex Gaussian noise), (iii) similar to the previous property, $\mathbb{E}(Z \varepsilon_{ic}^{[l]}(t))=\mathbb{E}(Z)\mathbb{E}(\varepsilon_{ic}^{[l]}(t))=0$ for all $Z$ independent of $\varepsilon_{ic}^{[l]}(t)$ (zero mean complex white Gaussian noise) and finally (iv) $g_{kc}^{[l]}$ and $\tilde g_{k'c'}^{[l]}$ are independent for all $(k,c)\neq(k',c')$. We now compute the four terms in Equation $(27)$. The last term, i.e., 
				\begin{align}\tag{28}
				\mathbb{E}(|W_u(t)/\sqrt{P_u}|^2)= \frac{1}{P_u}. 
				\end{align}
				Next we compute the second term in Equation $(27)$, namely,
				\begin{align}\tag{29}
				&\mathbb{E}(\sum_{c=1}^C |(\rho_{gc}^{[l]})^{d_g}\tilde{g}_{gc}^{[l]}(t-d_g)|^2)= \sum_{c=1}^C (\rho_{gc}^{[l]})^{2d_g} \left( \beta_{gc}^{[l]}-\frac{(\beta_{gc}^{[l]})^2}{\frac{1}{P_p}+\sum_{b=1}^C\beta_{gb}^{[l]}}     \right).
				\end{align}
				the latter is satisfied due to the fact that the variance of $\tilde g_{gc}^{[l]}(t-d_{g})$ being given by $\beta_{gc}^{[l]}-\frac{(\beta_{gc}^{[l]})^2}{\frac{1}{P_p}+\sum_{b=1}^C\beta_{gb}^{[l]}}$ for all $g$ and $c$.

				We now compute the third term in Equation $(27)$, that is,
				\begin{align}\tag{30}
				\nonumber\mathbb{E}\left(\sum_{c=1}^C\sum_{j=0}^{d_g-1} |\sqrt{\beta_{gc}^{[l]}}(\rho_{gc}^{[l]})^j\varepsilon_{gc}^{[l]}(t-j)|^2\right)&=\sum_{c=1}^C\sum_{j=0}^{d_g-1} \beta_{gc}^{[l]}(\rho_{gc}^{[l]})^{2j}   (1-(\rho_{gc}^{[l]})^{2} )\nonumber\\
				\nonumber&=\sum_{c=1}^C \beta_{gc}^{[l]} \frac{1-(\rho_{gc}^{[l]})^{2d_g}}{1-(\rho_{gc}^{[l]})^{2}}
				(1-(\rho_{gc}^{[l]})^{2})\nonumber\\
				\nonumber&=\sum_{c=1}^C \beta_{gc}^{[l]} ({1-(\rho_{gc}^{[l]})^{2d_g}}),\nonumber\\\nonumber
				\end{align}
				for the second equality we have used the expression of finite geometric sums since $(\rho_{gc}^{[l]})^2<1$ for all $g$ and $c$.
				We are left with the first term in Equation $(27)$, that is,
				\begin{align}\tag{31}
				&\sum_{k\neq g}^{N_g}\sum_{c=1}^C\mathbb{E}(|g_{kc}^{[l]}(t)|^2)=\sum_{k\neq g}^{N_g}\sum_{c=1}^C\mathbb{E}(|\sqrt{\beta_{kc}^{[l]}}     h_{kc}^{[l]}(t)|^2)=\sum_{k\neq g}^{N_g}\sum_{c=1}^C\beta_{kc}^{[l]},\nonumber\\\nonumber
				\end{align}
				Combining all four terms, that is, Equations $(28)$, $(29)$, $(30)$ and $(31)$, we obtain
				\begin{align}\tag{32}
				\E{|u_{gl}^\dagger(t) I_3(t)|^2}&=\sum_{k\neq g}^{N_g}\sum_{c=1}^C \beta_{kn}^{[l]} +
				\sum_{c=1}^C \beta_{gc}^{[l]} ({1-(\rho_{gc}^{[l]})^{2d_g}})
				+\sum_{c=1}^C (\rho_{gc}^{[l]})^{2d_g} \left( \beta_{gc}^{[l]}-\frac{(\beta_{gc}^{[l]})^2}{\frac{1}{P_p}+\sum_{b=1}^C\beta_{gb}^{[l]}}     \right)\nonumber\\
				&=\sum_{k\neq g}^{N_g}\sum_{c=1}^C \beta_{kc}^{[l]} + \sum_{ c=1}^{C} ( \beta^{[l]}_{gc} -    \rho^{[l]^{2 d_g}}_{ gc }\frac{ \beta^{[l]^2}_{gc}}{\frac{1}{P_{p}}+\sum_{b=1}^{C}  \beta^{[l]}_{gb} })+\frac{1}{P_u}.\nonumber\\\nonumber
				\end{align}
				
				Substituting the results in Equations $(24)$, $(26)$ and $(32)$ in Equation $(21)$, we obtain
				\begin{align}\tag{33}
				\overline R_{g,l}^0& \geq \left(1-\frac{\tau}{T_s}\right)\log\left(1+\frac{(\rho_{gl}^{[l]})^{2d_{g}}|u_{gl}^\dagger(t) \hat g_{gl}^{[l]}(t-d_{g})|^2}{F}\right),
				\end{align} 
				with
				\begin{align}\tag{34}
				F=&|u_{gl}^\dagger(t) \hat g_{gl}^{[l]}(t-d_{g})|^2     \sum_{c \neq l}^C (\rho^{[l]}_{ gc })^{2 d_{g}} \frac{\beta^{[l]^2}_{gc}}{\beta^{[l]^2}_{gl}}+\sum_{k\neq g}^{N_g}\sum_{c=1}^C \beta_{kc}^{[l]} + \sum_{ c=1}^{C} ( \beta^{[l]}_{gc} -    \rho^{[l]^{2 d_g}}_{ gc }\frac{ \beta^{[l]^2}_{gc}}{\frac{1}{P_{p}}+\sum_{b=1}^{C}  \beta^{[l]}_{gb} })+\frac{1}{P_u}.\nonumber
				\end{align}
				From Equation $(20)$ and Equation $(33)$ we obtain
				\begin{align}\tag{35}
				\overline R_{g,l}=\mathbb{E}(\overline R_{g,l}^0)& = \mathbb{E}\left(\left(1-\frac{\tau}{T_s}\right)\log\left(1+\frac{(\rho_{il}^{[l]})^{2d_{g}}|u_{il}^\dagger(t) \hat g_{il}^{[l]}(t-d_{g})|^2}{F}\right)\right)\nonumber\\
				&=\mathbb{E}\left(\left(1-\frac{\tau}{T_s}\right)\log\left(1+\frac{(\rho_{gl}^{[l]})^{2d_{g}}}{G}\right)\right),\nonumber
				\end{align} 
				where
				\begin{align*}
				G=&\frac{F}{|u_{gl}^\dagger(t) \hat g_{gl}^{[l]}(t-d_{g})|^2}.
				\end{align*}
				In order to compute the final expression of the bound on the average rate, it now suffices to compute the explicit expression of the right hand side (RHS) of Equation $(35)$. In order to do so, we apply Jensen's inequality to the RHS in Eq. $(35)$, that is,  
				\begin{align}\tag{36}
				\overline R_{g,l}&\geq\left(1-\frac{\tau}{T_s}\right)\log\left(1+\frac{(\rho_{g,l}^{[l]})^{2d_{g}}}{\mathbb{E}(G)}\right),
				\end{align}
				with
				\begin{align*}\tag{37}
				\mathbb{E}(G)=& \sum_{c \neq l}^C (\rho^{[l]}_{ gc })^{2 d_{g}} \frac{\beta^{[l]^2}_{gc}}{\beta^{[l]^2}_{gl}}+ \mathbb{E}\left(\frac{1}{|u_{gl}^\dagger(t) \hat g_{gl}^{[l]}(t-d_{g})|^2 }\right)\\
				&\cdot\left(\sum_{k\neq g}^{N_g}\sum_{c=1}^C \beta_{kc}^{[l]} + \sum_{ c=1}^{C} ( \beta^{[l]}_{gc} -    \rho^{[l]^{2 d_g}}_{ gc }\frac{ \beta^{[l]^2}_{gc}}{\frac{1}{P_{p}}+\sum_{b=1}^{C}  \beta^{[l]}_{gb} })+\frac{1}{P_u}\right).
				\end{align*}
				Note that $\left| u_{gl}^\dagger(t)  \hat{g}^{[l]}_{ gl } (t-d_{g})\right|^2$ has a Gamma distribution with parameters $(M,\frac{\beta^{[l]^2}_{gl}}{\frac{1}{P_p}+\sum_{b=1}^{C} \beta^{[l]}_{gb} })$. Consequently, the mean value of $\frac{1}{\left|u_{gl}^\dagger(t)  \hat{g}^{[l]}_{ gl } (t-d_{g})\right|^2}$ (that has an inverse Gamma distribution) is equal to $\frac{1}{(M-1) \times \frac{\beta^{[l]^2}_{gl}}{\frac{1}{P_p}+\sum_{b=1}^{C} \beta^{[l]}_{gb} }}$. Combining this together with the results in Equations $(37)$ and $(35)$ we obtain the desired lower bound on the average achievable spectral efficiency of user $g,l$, that is,
				\begin{align*}\tag{38}
				\overline R_{g,l} &  
				\geq \left( 1-\frac{\tau}{T_s}\right)  \text{log} \left(1+ \frac{(M-1)(\beta^{[l]}_{gl})^2 (\rho^{[l]}_{ gl })^{2d_g} }{(M-1) I^p_{gl} +  I^n_{gl}  } \right),
				\end{align*}
				where $I^p_{gl}$ and   $I^n_{gl}$ are given by
				\begin{align*}
				& I^p_{gl} = \sum_{ c \neq l}^{C} (\rho^{[l]}_{ gc })^{2d_{g}} (\beta^{[l]}_{gc})^2, \hbox{ and } \\
				& I^n_{gl}=  (\frac{1}{P_p}+\sum_{b=1}^{C} \beta^{[l]}_{gb}  )\cdot\left(\sum_{k\neq g}^{N_g}\sum_{c=1}^C \beta_{kc}^{[l]} + \sum_{ c=1}^{C} ( \beta^{[l]}_{gc} -    \rho^{[l]^{2 d_g}}_{ gc }\frac{ \beta^{[l]^2}_{gc}}{\frac{1}{P_{p}}+\sum_{b=1}^{C}  \beta^{[l]}_{gb} })+\frac{1}{P_u}\right).
				\end{align*}	
				Summing the achievable spectral efficiency of all grouped users concludes the proof.

			\subsection{Proof of Theorem 2}

		In order to  prove theorem 2, we consider the assymptotic regime where the number of  BS antennas $M$ grows very large. In this  case the lower bound on  achievable spectral efficiency of each user $g,l$ converges to the  following limit:
		
		\begin{align}\tag{39}
		\nonumber&  
		\left( 1-\frac{\tau}{T_s}\right) \text{log} \left(1+ \frac{(M-1)\beta^{[l]^2}_{gl} \rho^{[l]^{2 d_g}}_{ gl } }{(M-1) \times I^p_{gl} +  I^n_{gl}  } \right) \longrightarrow 
		\left( 1-\frac{\tau}{T_s}\right) \text{log} \left(1+ \frac{\beta^{[l]^2}_{gl} \rho^{[l]^{2 d_g}}_{ gl } }{\sum_{ b\neq l}^{C} \rho^{[l]^{2 d_g}}_{ gb } \beta^{[l]^2}_{gb} } \right).
		\end{align}
		The proposed framework is compared with  a reference  massive MIMO system  where, all scheduled users are required to perform uplink  training. In the asymptotic regime, the lower bound on the  achievable spectral efficiency of each user $g,l$  in the reference system  converges to the following limit:
		\begin{align}\tag{40}
		\left( 1-\frac{N_g}{T_s}\right) \text{log} \left(1+ \frac{\beta^{[l]^2}_{gl} }{\sum_{ b\neq l}^{C}  \beta^{[l]^2}_{gb} } \right).
		\end{align}
		The aim here, is to improve the  achievable spectral efficiency of each scheduled users. Consequently, the  spectral efficiency  of each user in the two considered systems should verify, $\; \forall\; g=1...N_g, l=1...C$:
		\begin{align}\tag{41}
		\nonumber& \left( 1-\frac{\tau}{T_s}\right) \text{log} \left(1+ \frac{\beta^{[l]^2}_{gl} \rho^{[l]^{2 d_g}}_{ gl } }{\sum_{ b\neq l}^{C} \rho^{[l]^{2 d_g}}_{ gb } \beta^{[l]^2}_{gb} } \right) \geq  \left( 1-\frac{N_g}{T_s}\right) \text{log} \left(1+ \frac{\beta^{[l]^2}_{gl} }{\sum_{ b\neq l}^{C}  \beta^{[l]^2}_{gb} } \right).
		\end{align}
		$(27)$ is equivalent  to the following condition:
		\begin{align}\tag{42}
		\nonumber&  \frac{\beta^{[l]^2}_{gl} \rho^{[l]^{2 d_g}}_{ gl } }{\sum_{ b\neq l}^{C} \rho^{[l]^{2 d_g}}_{gb } \beta^{[l]^2}_{gb} } \geq  \left(1+\frac{\beta^{[l]^2}_{gl} }{\sum_{ b\neq l}^{C}  \beta^{[l]^2}_{gb} }\right)^{\frac{T_s-N_g}{T_s-\tau}}-1.
		\end{align}
		We consider the extreme case where $\rho^2_{gl}=\bar{\rho}^{[{min}]^2}_{g}$ and $\forall b\neq l, \rho^2_{gb} =\bar{\rho}^{[{max}]^2}_{{g}}$. 	Here $\bar{\rho}^{[{min}]}_{g}$ and $\bar{\rho}^{[{max}]}_{{g}}$ denote respectively the minimum and maximum channel autocorrelation  coefficients in  group $g$. This means that we assume the worst case scenario for each user where, its coherence time is always lower than its fellow copilot users.
		Finally, by considering $SINR^{[\infty]}_{g,l}= \frac{\beta^{[l]^2}_{gl} }{\sum_{ b\neq l}^{C}  \beta^{[l]^2}_{gb} }$, we obtain $(11)$ which finishes the proof.

	\subsection{Proof of Theorem 3}  
		In order to prove theorem $3$, we start by demonstrating that the objective function $f$ in $(14)$, is submodular.
		We consider  two copilot groups  allocations, $X \; \text{and} \; Z$ such that  $X \subseteq Z$. We  define the following ground set  $G=\left\lbrace x^{[g]}, g=1...N_g  \right\rbrace$, where $x^{[g]}, g=1...N_g$ is an abstract element  denoting the scheduling of copilot group $g$ for uplink training. We need to prove that the marginal value of scheduling a new copilot group $x^{[n]}$  in $X$ and $Z$ verifies:
		\begin{align}\tag{43}
		\nonumber& f\left(X \cup \left\lbrace x^{[n]}\right\rbrace \right) - f\left(X  \right) \geq f\left(Z \cup \left\lbrace x^{[n]}\right\rbrace \right) - f\left(Z  \right).\nonumber
		\end{align}	
		We start by  computing  the marginal values of adding an element $x^{[n]}$ to $X$ and  $Z$ which are, respectively, given by: 
		\begin{align}\tag{44}
		\nonumber& f\left(X \cup \left\lbrace x^{[n]}\right\rbrace \right) - f\left(X  \right) = \left( \frac{\lvert X \rvert}{T_s}-\frac{\lvert X \rvert+1}{T_s}\right) \{\sum_{l=1}^{C} \sum_{x^{[i]} \in X}^{} w_{il}\text{log} \left(1+ \frac{(M-1)\beta^{[l]^2}_{il} }{(M-1) \times I^p_{il} +  I^n_{il}  } \right)\\
		\nonumber& + \sum_{x^{[i]} \in G \backslash [X\cup x^{[n]}] }^{} w_{il} \text{log} \left(1+ \frac{(M-1)\beta^{[l]^2}_{il} \rho^{[l]^{2 d_i}}_{ il } }{(M-1) \times I^p_{il} +  I^n_{il}  } \right)\}  + (1-\frac{\lvert X \rvert+1}{T_s})  w_{nl}\text{log} \left(1+ \frac{(M-1)\beta^{[l]^2}_{nl}  }{(M-1) \times I^p_{nl} +  I^n_{nl}  } \right)\\
		\nonumber& 	-(1-\frac{\lvert X \rvert}{T_s})w_{nl}\text{log} \left(1+ \frac{(M-1)\beta^{[l]^2}_{nl} \rho^{[l]^{2 d_n}}_{ nl } }{(M-1) \times I^p_{nl} +  I^n_{nl}  } \right),\\	\nonumber
		\end{align}	
		and,
		\begin{align}\tag{45}
		\nonumber& f\left(Z \cup \left\lbrace x^{[n]}\right\rbrace \right) - f\left(Z  \right) = \left(\frac{\lvert Z \rvert}{T_s} -\frac{\lvert Z \rvert+1}{T_s}\right)  \{\sum_{l=1}^{C} \sum_{x^{[i]} \in Z}^{} w_{il}\text{log} \left(1+ \frac{(M-1)\beta^{[l]^2}_{il} }{(M-1) \times I^p_{il} +  I^n_{il}  } \right)\\
		\nonumber& + \sum_{x^{[i]} \in G \backslash [Z\cup x^{[n]}] }^{} w_{il}\text{log} \left(1+ \frac{(M-1)\beta^{[l]^2}_{il} \rho^{[l]^{2 d_i}}_{ il } }{(M-1) \times I^p_{il} +  I^n_{il}  } \right) \}+ (1-\frac{\lvert Z \rvert+1}{T_s})w_{nl}\text{log} \left(1+ \frac{(M-1)\beta^{[l]^2}_{nl}  }{(M-1) \times I^p_{nl} +  I^n_{nl}  } \right)\\
		\nonumber& 	-(1-\frac{\lvert Z \rvert}{T_s})w_{nl}\text{log} \left(1+ \frac{(M-1)\beta^{[l]^2}_{nl} \rho^{[l]^{2 d_n}}_{ nl } }{(M-1) \times I^p_{nl} +  I^n_{nl}  } \right).	\nonumber
		\end{align}	
		The difference between the two marginal  values is given by:
		\begin{align}\tag{46}
		\nonumber& f\left(X \cup \left\lbrace x^{[n]}\right\rbrace \right) - f\left(X  \right) - (f\left(Z \cup \left\lbrace x^{[n]}\right\rbrace \right) - f\left(Z  \right))=\\
		\nonumber& [\frac{\lvert Z \rvert-\lvert X\rvert}{T_s}] w_{nl}\text{log} \left(1+ \frac{(M-1)\beta^{[l]^2}_{nl}  }{(M-1) \times I^p_{nl} +  I^n_{nl}  } \right)
		+[\frac{\lvert X \rvert-\lvert Z\rvert}{T_s}]w_{nl} \text{log} \left(1+ \frac{(M-1)\beta^{[l]^2}_{nl} \rho^{[l]^{2 d_n}}_{ nl } }{(M-1) \times I^p_{nl} +  I^n_{nl}  } \right)\\
		\nonumber& + \sum_{l=1}^{C} \sum_{x^{[i]} \in Z \backslash X}^{} [\frac{w_{il}}{T_s}] \left( \text{log} \left(1+ \frac{(M-1)\beta^{[l]^2}_{il} }{(M-1) \times I^p_{il} +  I^n_{il}  } \right )   -  \text{log} \left(1+ \frac{(M-1)\beta^{[l]^2}_{il} \rho^{[l]^{2 d_i}}_{ il } }{(M-1) \times I^p_{il} +  I^n_{il}  } \right )  \right). \nonumber
		\end{align}		
		
		The difference  in  marginal values is positive. Consequently, the objective function $f$ is submodular. In addition, it is clear that  $(14a)$ is a cardinality constraint.
		Problem $(14)$ is then  equivalent  to  maximizing a submodular function with a cardinality constraint.

\end{document}